\documentstyle[preprint,aps]{revtex}

\begin{document}

\draft
\preprint{PSU/TH/159}
\title{ Warm Inflation}
\author{Arjun Berera}
\address{Department of Physics, The Pennsylvania State University, University
Park, Pennsylvania, 16802, USA}
\date{\today}
\maketitle

\begin{abstract}
Assuming a first order phase transition during inflation,
a model scenario is described which does not
require a tiny
coupling constant.  Thermal equilibrium is closely maintained
as inflation commences.  No large scale reheating is necessary.
Solutions for $N> 70$ are found for any $\delta \rho / \rho$ in the range
$10^{-5} - 10^{-3}$.

\medskip

In press Physical Review Letters 1995

\medskip
\end{abstract}

\narrowtext
The standard slow-roll inflation model separates expansion
and reheating into two distinguished time periods.
It is first assumed that exponential expansion
from inflation places the universe in a super-cooled phase.
Subsequently thereafter the universe is reheated.
Two outcome arise from such a
scenario.  Firstly the required density perturbations
in this cold universe are left to be created by
the quantum fluctuations of the inflaton.
Secondly, the temperature cliff after expansion
requires a temporally localized mechanism that
rapidly distributes sufficient vacuum energy for reheating.

It was recognized by Berera and Fang \cite{bf1}
(hereafter referred to as (I) )  that meshing these two isolated
stages may resolve the disparities created by each separately.
In (I) it was shown that slow-roll inflation
\cite{kolb,brand}
is parametrically consistent with a thermal component.
The thermodynamic analysis there made plausible, but gave no
explicit model scenario.
In this paper
we present a model example in which thermal equilibrium
is maintained during the inflation stage.
The result depends
on the presence of a first order phase transition.  In such a
case, below $T_c$ there is a temperature barrier which separates
the symmetry broken and unbroken phases.  Near
$T_c$, the region
from the top of the
temperature barrier to the symmetry broken minima
is somewhat flat.   Furthermore, beyond the inflection point,
where the curvature is positive, thermal fluctuations
of the inflaton are damped.
It appears that these two conditions can support
a slow-roll solution for
unexceptional values of the coupling constant,
provided that there is also a dissipative
component of sufficient size in the inflaton's
equation of motion.

To demonstrate this, we will examine the
finite-temperature one-loop Coleman-Weinberg
effective potential in the form \cite{amit,linde},
\begin{equation}
V (\phi, T) = {1 \over 2} B M^4 +
B\phi^4[{\ln}({{\phi^2} \over {M^2}}) -{1 \over 2}] + CT^2 \phi^2 .
\label{cwpot}
\end{equation}
At temperatures greater than the critical temperature $T_c$, the
effective potential has a unique minima at $<\phi> = 0$.
For $0<T<T_c$, there are two minima separated by a temperature barrier.
One is a metastable minima at $<\phi> = 0$ and the
other lies at $<\phi>  \equiv \phi_B \sim M$.  These valleys are
separated by a hill with the peak position defined
as $\phi_T$.  Hereafter, we will set the GUT scale $M=1$ unless otherwise
specified.

In (I) we had found the equation of motion
of the inflaton with also a dissipative term to be,
\begin{equation}
<(3H+\Gamma (\phi)) \dot{\phi} + V'(\phi,T)>=0
\label{eom}
\end{equation}
where the Hubble constant
\begin{equation}
H = M({M \over {m_{pl}}}) \sqrt{ {{8\pi V(0)} \over {3M^4}} }
\end{equation}
with $m_{pl} =1.2 \times 10^{19} GeV$ and
in the model of eq. (\ref{cwpot}) $V(0)=BM^4/2$.
Here $<\phi>$ means the thermal average of the inflaton field $\phi$.
In the region of interest we will
find $< \phi> \gg \delta \phi$.  Thus we can remove
the averaging brackets and treat eq. (\ref{eom}) classically.
We will also denote $\Gamma_{\phi} \equiv <\Gamma (\phi)>$,
and $\Gamma$ as the average with respect to
$dN/d\phi$ of $\Gamma_{\phi}$
over the slow-roll interval at fixed T.

The slow-roll conditions
\begin{equation}
\dot{\phi}^2/2 \ll V(\phi,T)
\end{equation}
and
\begin{equation}
\ddot{\phi} \ll (3H+ \Gamma_{\phi}) \dot{\phi}
\end{equation}
will require respectively that
\begin{equation}
| V'(\phi,T) | \ll \sqrt{2 V(\phi, T)} (3H + \Gamma_{\phi})
\label{vp}
\end{equation}
and
\begin{equation}
| V''(\phi,T) | \ll (3H+\Gamma_{\phi})^2.
\label{vpp}
\end{equation}
When there is a sizable thermal component, eq. (\ref{vp})
is replaced by the condition,
\begin{equation}
R \equiv {{{1 \over 2} \dot{\phi}^2 + \rho_r} \over {V(\phi,T)}}
\ll 1.
\end{equation}
For later use,  near the critical
point, for $T<T_c$ in the region about $\phi_B$,
the effective potential behaves as,
\begin{equation}
V(\phi,T) \approx c_1 T^2 (\phi -\phi_B)^2
\label{apppot}
\end{equation}
where $c_1$ and $c_2$ have logarithmic dependence on T.
%It will be convenient in some cases to factor out the $T^2$ dependence
%by defining
%\begin{equation}
%\tilde{V}(\phi) \equiv {{V(\phi)} \over {T^2}}.
%\end{equation}
%In the model
%studied below they are of order 1.

Within the slow-roll regime, the number of e-folds from
$\phi_T < \phi_i$ to $\phi_i < \phi_f < \phi_B$ is
\begin{equation}
N=H \int_{\phi_i}^{\phi_f} d\phi
{{3H+\Gamma_{\phi}} \over {-V'(\phi,T)}}.
\label{n1}
\end{equation}
The amplitude of density perturbation \cite{kolb,brand,peebl}
at the onset of slow-roll,
when there is also thermal radiation, is from
(I-25),
\begin{equation}
\Delta \equiv {{\delta \rho} \over {\rho}}=
{{-\delta \phi V'(\phi,T)} \over {\dot{\phi}^2+{4 \over 3}\rho_r}}
\end{equation}
where following (I), for the region $V'' \gg H^2$,
\begin{equation}
\delta \phi = \sqrt{ {3 \over {4\pi}} H T \frac{H^2}{V''(\phi, T)}}.
\end{equation}
Using
\begin{equation}
\rho_r \simeq {{\Gamma_{\phi}} \over {4 H}} \dot{\phi}^2
\label{rrho}
\end{equation}
from (I-9), and
\begin{equation}
\rho_r ={{ \pi^2 } \over {30}} g^* T^4
\end{equation}
one obtains
\begin{equation}
\Delta(\phi) = {9 \over 4} \sqrt{{10 \over {\pi^3}}} {1 \over {\sqrt{g^*}}}
({{\Gamma_{\phi}} \over H})^{1 \over 2} ({H \over T})^{3 \over 2}
\left(\frac{H^2}{V''(\phi, T)}\right)^{\frac{1}{2}}
\label{del2}
\end{equation}
and the auxiliary condition,
\begin{equation}
\sqrt{{15} \over {2 \pi^2}} \sqrt{{\Gamma_{\phi}} \over {g^*H}}
{{|V'(\phi,T)|} \over {T^2(3H+\Gamma_{\phi}})} =1.
\label{aux}
\end{equation}
Here $g^*$ is the effective number of degrees of freedom, which
implicitly depends on T.  When $T>M_Z$ (ie. the mass of the Z particle),
$g^* =423/4$.
By setting $T \sim 0.1 M$ in eq. (\ref{apppot})
in order
to satisfy eq. (\ref{vpp}) it will
require
\begin{equation}
\Gamma_{\phi} \gg H.
\end{equation}

In this limit,
substituting for $\Gamma_{\phi}$
from eq. (\ref{aux}) into eqs. (\ref{n1}) and (\ref{del2})
we get,
\begin{equation}
N =-\frac{15}{2\pi^2} \frac{1}{g^*T^4} \int_{\phi_i}^{\phi_f}
V'(\phi,T)d \phi
\label{n2}
\end{equation}
and
\begin{equation}
\Delta_{\phi} \equiv
\Delta(\phi) = \frac{45}{4} \sqrt{\frac{3}{\pi^5}}
\frac{1}{g^*} \left(\frac{H}{T}\right)^{\frac{3}{2}}
\frac{-V'(\phi,T)}{T^2\sqrt{V''(\phi,T)}}
\end{equation}
For later,
the notation $\Delta$ with no
subscript will refer to $\Delta_{\phi}$
averaged with respect to $dN/d\phi$ over
the slow-roll region.
The slow-roll motion here is likened to viscous damping
by $\Gamma_{\phi}$ along a mildly cusped
potential surface.

For our numerical study, we will use the values from
the SU(5) Georgi-Glashow model \cite{gegl} in eq.
(\ref{cwpot})
of \cite{linde},
\begin{equation}
B= {5625 \over {1024 \pi^2}} g^4
\label{bb}
\end{equation}
and
\begin{equation}
C= {75 \over {16}}  g^2.
\label{cc}
\end{equation}
The acceptable SU(5) range for the coupling
constant is $0.5 <$ g $<0.6$.  In what follows, we will
set g=0.57.

In this model, the following scaling relations are satisfied,
\begin{equation}
N(xg,xT)=N(g,T),
\label{scale}
\end{equation}
\begin{equation}
\Delta_{\phi}(xg,xT)=x^{\frac{3}{2}} \Delta_{\phi}(g,T).
\label{scaled}
\end{equation}
and
\begin{equation}
\Gamma_{\phi}(xg,xT)=x^2\Gamma_{\phi}(g,T).
\label{scaled2}
\end{equation}
The relations (\ref{scaled}) and (\ref{scaled2}) also
hold for $\Delta$ and $\Gamma$ respectively.
With respect to the model,
\begin{equation}
R_0 \equiv R(\phi=0) = {{\pi^2} \over {15}} {{g^*T^4} \over {BM^4}}
\end{equation}
where $B$ is given by eq. (\ref{bb}) and
where we have explicitly written the M dependence.
Since
\begin{equation}
R_0(xg,xT)=R_0(g,T)
\end{equation}
and
\begin{equation}
T_c(xg)=xT_c(g),
\end{equation}
the critical point is an invariant point $R_c$ in $R_0$ for all g.
For the parameters (\ref{bb}) and (\ref{cc}), we find
$R_c \approx 0.65$ and $T_c(g=0.57)=0.15$.

By holding g fixed and ranging over T, one obtains a single universal
curve from which $N$ can be computed for any choice of g and T.
With this, the model is completely solved for $N$.
It is not possible to arbitrarily adjust the coupling constant
in this model, since
varying g also alters the thermodynamic properties
of the system.  From eq. (\ref{scale}) it
is equivalent
to holding g fixed and varying T.
As such, in this thermal scenario, "fine tuning"
has meaning of thermodynamics.  The central problem
becomes the common one of determining the behavior
of the system with respect to its intensive variable, here
temperature.

In fig. (1) we have computed $N$
for g$= 0.57$ and $M=10^{15}$ GeV
in the region $V''>0$
and $V'<0$ where eqs (\ref{vp}) and (\ref{vpp})
hold.  In the same figure, the dashed curve is
$\alpha \Gamma$ with $\alpha = 2.5 \times 10^{-4}$.
In fig. (2), $\Delta$ is given over the same
range of temperatures.

{}From fig. (1) one finds that
solutions consistent with observation are for $T<0.033$.
For example at T=0.032 where $R_0=1.2 \times 10^{-3}$,
we find N=73, $\Delta \simeq 10^{-4}$
and $\Gamma \simeq 3.6 \times 10^5$.
There is a monotonic decrease in $\Delta_{\phi}$ and $\Gamma_{\phi}$
along the slow-roll trajectory.  Their variation is due to the respective
decrease and increase of $|V'(\phi,T)|$ and $V''(\phi,T)$
in this region, which is a common feature of free energy
functions that describe
first order transitions.   However, a factor 2-3
variation in $\Delta_{\phi}$ is within
observational uncertainties \cite{kolb,turste}.

Let us get an idea of how much $\Delta_{\phi}$ and $\Gamma_{\phi}$
change in the above case. By
expressing $\Delta_{\phi}$ and $\Gamma_{\phi}$
as functions of N rather than $\phi$,
we find that $\Delta(70)=3 \times 10^{-4}$,
$\Delta(1) = 0.8 \times 10^{-5}$ whereas $\Gamma_{\phi}$
varies from $5 \times 10^5$ to $1.6 \times 10^4$ between
the same limits.  However $\Gamma_{\phi}$ changes by
less than $20\%$ for the first 30 e-folds of inflation.

One can decrease these variations significantly by going to lower
temperature.  Here, the observational constraints can be
adequately satisfied within a region near the inflection
point, where $V'$ and $V''$ vary the least.  For
example at $T=0.02$ for about a quarter of the
slow-roll interval, $N>70$ can be attained with less than a $30\%$
variation in $\Delta_{\phi}$.  As one case, from $\phi_i=0.78$
to $\phi_f=0.81$, 72 e-folds are generated with
$\Delta(\phi_i) =4.9 \times 10^{-4}$,
$\Delta(\phi_f)=3.8 \times 10^{-4}$,
$\Gamma_{\phi}(\phi_i)=3.2 \times 10^6$
and $\Gamma_{\phi}(\phi_f)=2.9 \times 10^6$.
Going even lower in temperature to $T=0.01$, solutions
for $N>70$ lie within a small interval of
$\Delta \phi \equiv \phi_f - \phi_i$.  In this case,
the variation in $\Delta_{\phi}$ and $\Gamma_{\phi}$ is reduced to less than
$1\%$ with $\Delta \simeq 10^{-3}$ and $\Gamma \simeq 10^7$.

Although it is interesting to study the solutions contained
in this model from figures (1) and (2) and
the relations (\ref{scale})-(\ref{scaled2}),
since g$\phi / T > 1$, this model only semi-quantitatively
represents the finite temperature effects of the
SU(5) theory
\cite{dolan}.    In the relevant region which is $T \sim T_c/5$,
the free energy function that must be computed is at the lower
edge of the sensitive critical region.  Noting from
eq. (\ref{n2}) that N only depends on the end points of the
slow-roll interval, one expects less significant changes of it
in an improved treatment of temperature effects.
However the curvature $V''(\phi, T)$ and so the variation in
$\Delta_{\phi}$ and $\Gamma_{\phi}$
could be sizably different
in a careful treatment
of the exact finite temperature SU(5) theory.

In this paper we have formulated a simple-minded thermal scenario,
which already agrees reasonably well with observational
bounds \cite{smoot,benn}.  There are two immediate concerns
with our solution.  The first is the variation in the
"constant"  $\Gamma_{\phi}$ which changes by at most
a factor 10 for reasonable scenarios.  When interpreted in terms
of temperature from eq. (\ref{rrho}), this would imply
a factor 1.7 variation in T.  One could eliminate this
variation by including the $\dot \rho_r$ term in eq. (\ref{rrho})
and forego the limit of perfect equilibrium.
Equilibrium results
within the above range of temperature remain
in the set of observationally consistent solutions
for N and $\Delta$.  Using this as a guide, one expects
this to give some estimate when nonequilibrium dynamics is
treated.  One possibility for treating
nonequilibrium dynamics for the inflaton would be
with a modified version of the KPZ-equation \cite{kpz,bf2}.
However, at present we do not have a quantitative description
of the dynamics for such a situation.  The similar
problem of treating non-equilibrium dynamics
also arises for reheating in the standard inflation scenario.
To our knowledge, there is no satisfactory resolution to this problem
there either.  In the present case, since equilibrium estimates
are already promising, this problem
now has primary importance in formulating a complete
thermal scenario.

The second concern is in regards to the magnitude of $\Gamma_{\phi}$.
We can get some particle physics
estimates by treating $\phi(x)$ as an external source
for particle creation.  For example, consider the Yukawa interaction,
\begin{equation}
L_I = g_y \phi(x) \bar{\psi}(x) \psi(x)
\end{equation}
where the fermions are light with mass $\mu$.
The probability P to create a fermion pair is \cite{Izuber},
\begin{equation}
P= \frac{g_y^2}{8\pi^2} \int \frac{d^4p}{(2\pi)^4}
|\tilde{\phi}(p)|^2 p^2(1-\frac{4\mu^2}{p^2})^{3/2}
\theta(p^2-4\mu^2)
\end{equation}
where
\begin{equation}
\tilde{\phi}(p) = \int d^4x \phi(x) e^{ip\cdot x}
\end{equation}
with
\begin{equation}
\tilde{\phi}(p) \sim MVT
\end{equation}
for the lowest mode $p \sim H$.
Using $V \sim 1/H^3$,
the decay rate is,
\begin{equation}
\Gamma_y = \frac{P}{T} \sim g_y^2 \frac{M}{H} M,
\end{equation}
which is of the same scale as what we found for $\Gamma_{\phi}$.
Although this numerical correspondence is noteworthy,
at present there is no formal connection between
estimates such as $\Gamma_y$
and $\Gamma_{\phi}$.

To summarize, this paper has demonstrated a class of equilibrium
scenarios that parametrically satisfy the observational constraints.
To formulate a complete thermal scenario, the small variation
in $\Gamma_{\phi}$, which we believe is tied to nonequilibrium
dynamics, must be understood.

It is interesting to note an alternative interpretation of this
model scenario$^\dagger$.  The inclusion of thermal effects acts similar
to a mass term which breaks the scale symmetry of the
zero-temperature theory.  In \cite{ratra} it has been argued
that the coupling constant fine-tuning problem is closely
associated to this scale symmetry.  On general grounds breaking
this scale symmetry has been suggested there as a way to avoid
the small-coupling constant problem.  Accepting their argument,
the the thermal scenario presented here and more generally in (I)
can be regarded as an explicit realization of scale symmetry breaking.

Before concluding, three points should be clarified.  Firstly,
since the temperature in this scenario is near $T_c$, the process
of rapid cooling followed by rapid heating is replaced
by a smoothened dissipative mechanism. Secondly,
since
\begin{equation}
<\phi> \sim 30 T \gg \sqrt{HT} \gg \delta \phi,
\end{equation}
our
semiclassical treatment is
justified.  Finally in this thermal scenario there are no
issues about how quantum coherence is generated from an initially
random field configuration.
\bigskip

I thank Professor L. Z. Fang for many helpful suggestions.
Financial support was provided by
the U. S. Department of Energy, Division of High Energy and Nuclear
Physics.

\medskip

$^\dagger$ I thank the referee for this observation.

\eject
FIGURE CAPTIONS

figure 1: e-folds $N$ of inflation (solid line) and
$\alpha \Gamma$ (dashed line) versus
temperature with $\alpha = 2.5 \times 10^{-4}$,
and for
g=0.57
and $M=10^{15}$ GeV.

figure 2: $\Delta$ versus temperature with
g=0.57 and $M=10^{15}$GeV.
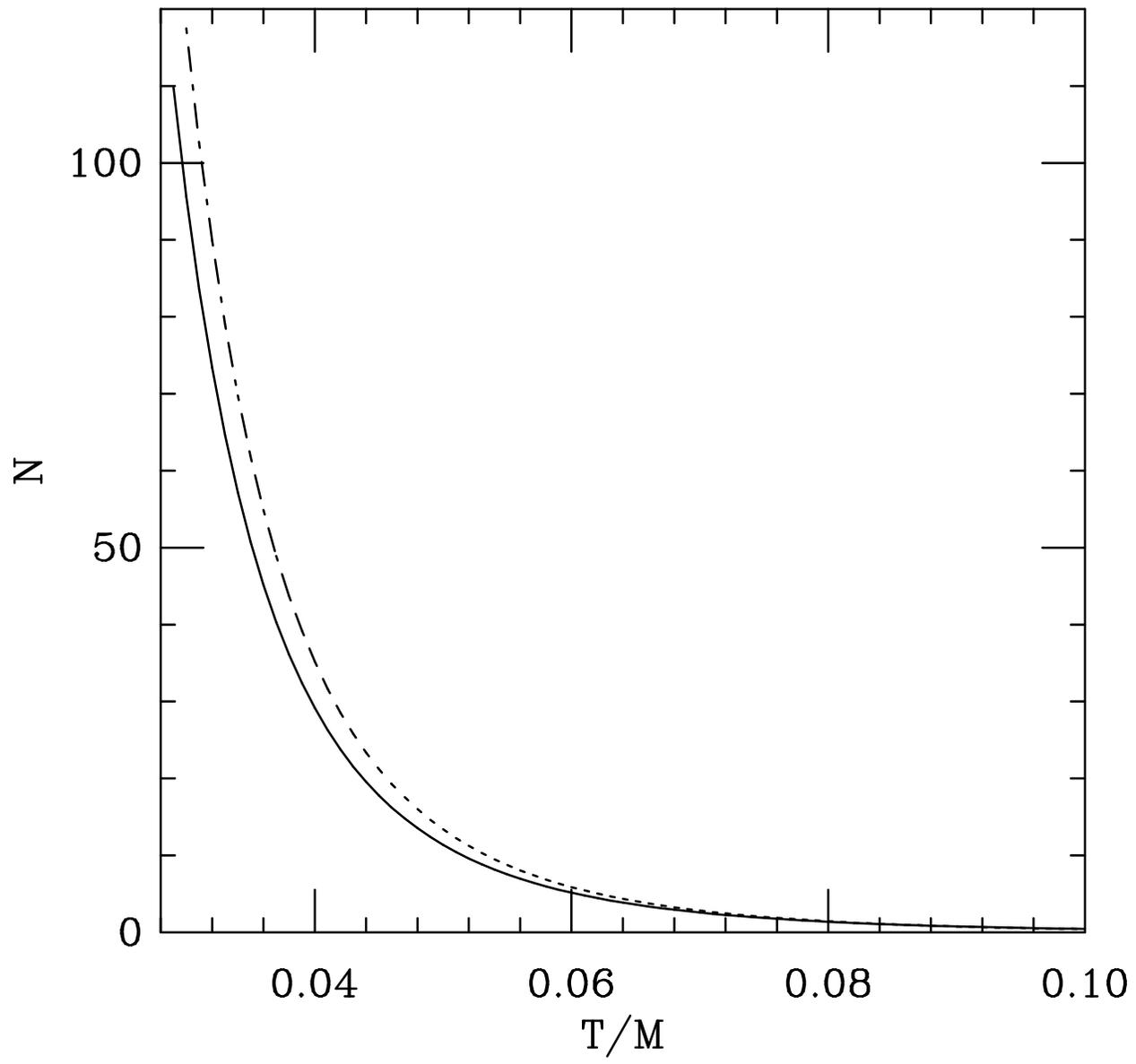
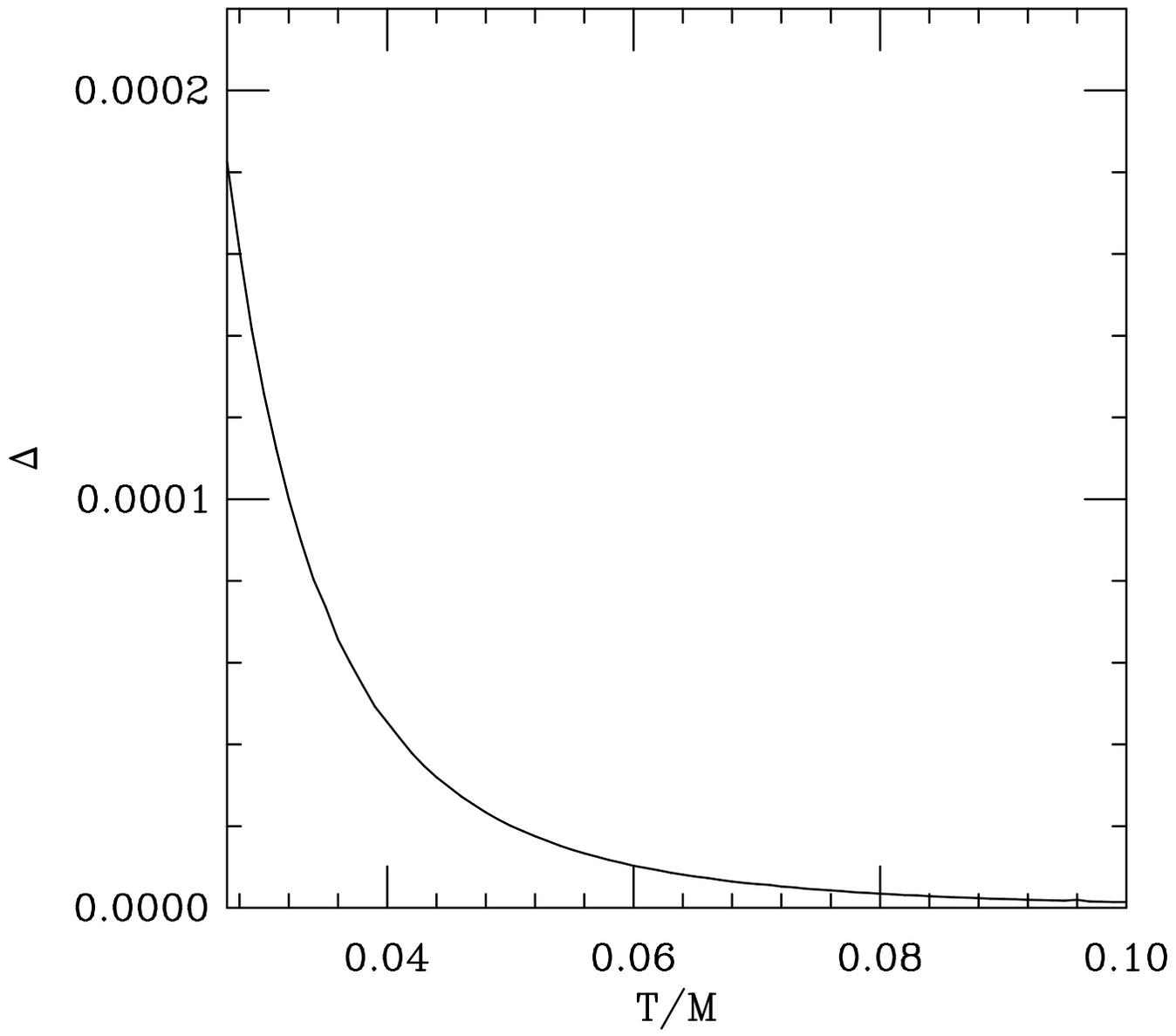
\end{document}